\documentclass[reprint, amsmath, amssymb, aps, prl, twocolumn, linenumbers, notitlepage, superscriptaddress]{revtex4-2}
\usepackage{upgreek}
\usepackage{graphicx}% Include figure files
\usepackage{dcolumn}% Align table columns on decimal point
\usepackage{bm}% bold math
\usepackage[colorlinks=true,
 linkcolor=blue,
 anchorcolor=blue,
 citecolor=blue,
 urlcolor=blue
 ]{hyperref}
\usepackage[dvipsnames]{xcolor}
\usepackage{physics}
\usepackage{braket}
\usepackage{siunitx}
\DeclareSIUnit\gauss{G}
\AtBeginDocument{\RenewCommandCopy\qty\SI}
\usepackage[utf8]{inputenc}

\makeatletter
\newcommand\appendixsection[1]{%
  \refstepcounter{section}%
  \vspace{-\baselineskip} 
  \subsection*{Appendix \Alph{section}: #1}%
  \addcontentsline{toc}{section}
  {Appendix \Alph{section}: #1}%
  \vspace{-\baselineskip} 
}
\makeatother

\begin{document}
\nolinenumbers

\title{Microwave electrometry with quantum-limited resolutions in a Rydberg-atom array}

\author{Yao-Wen Zhang}
\thanks{These authors contributed equally to this work.}
\author{De-Sheng Xiang}
\thanks{These authors contributed equally to this work.}
\author{Ren Liao}
\thanks{These authors contributed equally to this work.}
\author{Hao-Xiang Liu}
\thanks{These authors contributed equally to this work.}
\author{Biao Xu}
\thanks{These authors contributed equally to this work.}
\author{Peng Zhou}
\affiliation{
National Gravitation Laboratory, MOE Key Laboratory of Fundamental Physical Quantities Measurement,\\
Hubei Key Laboratory of Gravitation and Quantum Physics, PGMF,\\
Institute for Quantum Science and Engineering, School of Physics,\\
Huazhong University of Science and Technology, Wuhan 430074, China
}
\author{Yijia Zhou}
\affiliation{Shanghai Qi Zhi Institute, Shanghai 200232, China
}
\affiliation{Yangtze Delta Industrial Innovation Center of Quantum Science and Technology, Suzhou 215000, China}
\author{Kuan Zhang}
\email{zhang\_k09@hust.edu.cn}
\affiliation{
National Gravitation Laboratory, MOE Key Laboratory of Fundamental Physical Quantities Measurement,\\
Hubei Key Laboratory of Gravitation and Quantum Physics, PGMF,\\
Institute for Quantum Science and Engineering, School of Physics,\\
Huazhong University of Science and Technology, Wuhan 430074, China
}
\author{Lin Li}
\email{li\_lin@hust.edu.cn}
\affiliation{
National Gravitation Laboratory, MOE Key Laboratory of Fundamental Physical Quantities Measurement,\\
Hubei Key Laboratory of Gravitation and Quantum Physics, PGMF,\\
Institute for Quantum Science and Engineering, School of Physics,\\
Huazhong University of Science and Technology, Wuhan 430074, China
}
\affiliation{Wuhan Institute of Quantum Technology, Wuhan 430206, China}

\begin{abstract}
Microwave (MW) field sensing is foundational to modern technology, yet its evolution, reliant on classical antennas, is constrained by fundamental physical limits on field, temporal, and spatial resolutions.
Here, we demonstrate an MW electrometry that simultaneously surpasses these constraints by using individual Rydberg atoms in an optical tweezer array as coherent sensors. This approach achieves a field sensitivity within 13\% of the standard quantum limit, a response time that exceeds the Chu limit by more than 11 orders of magnitude, and in-situ near-field mapping with $\lambda/3000$ spatial resolution. This work establishes Rydberg-atom arrays as a powerful platform that unites quantum-limited sensitivity, nanosecond-scale response time, and sub-micrometer resolution, opening new avenues in quantum metrology and precision electromagnetic field imaging.
\end{abstract}
    
\maketitle

\begin{figure*}
    \centering
    \includegraphics[width=2\columnwidth]{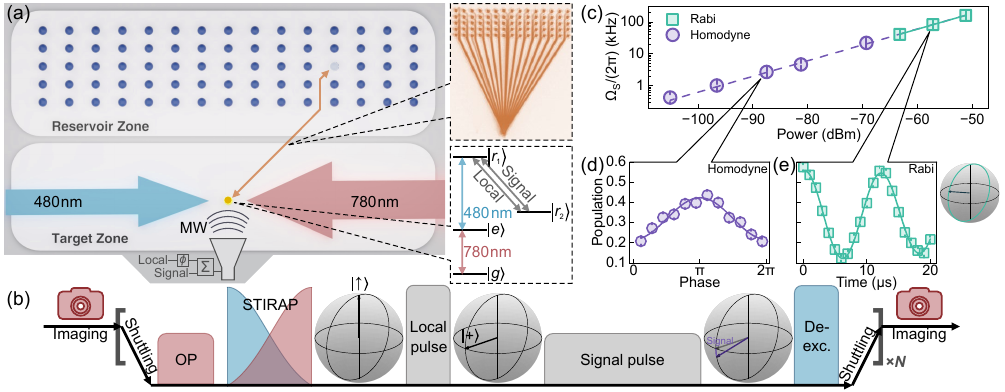}
    \caption{
    Microwave electrometry with Rydberg-atom array.
    (a) Schematic of the experimental setup. 
    The atoms are loaded and imaged in the reservoir zone, while the MW measurement is performed in the target zone. 
    A movable optical tweezer shuttles the atoms between the two zones. 
    The insets show the trajectory of the shuttling tweezer and the relevant atomic levels. 
    (b) Experimental sequence. 
    In the target zone, each atom undergoes optical pumping (OP), Rydberg excitation via STIRAP, MW measurement, and state-selective de-excitation. The local pulse is applied for weak-field measurements. See text and Appendix~\ref{APD:apparatus} for details.
    (c) Calibration of field-measurement schemes. 
    The weak-field single-atom homodyne measurement results of the Rabi frequency $\Omega_\mathrm{S}$ of the signal field (purple circles) agree well with the fitting curve obtained from strong-field Rabi-oscillation data (cyan squares). 
    (d) A representative single-atom homodyne measurement in the weak-field regime.
    (e) A typical Rabi oscillation in the strong-field regime.
    Error bars represent $1\sigma$ standard errors.
    }
    \label{Fig:Setup}
\end{figure*}

\emph{Introduction}---%
High-precision microwave (MW) sensing is critical for a wide spectrum of fields, from next-generation communications and medical imaging to deep-space exploration and scientific discovery.
However, the field, temporal, and spatial resolutions of classical antennas face fundamental limits: the sensitivity is bounded by the Johnson--Nyquist noise~\cite{johnson1928thermal,nyquist1928thermal}, the response time by the Chu limit (for a given antenna size)~\cite{chu1948physical}, and the spatial resolution is diffraction-limited to the MW wavelength scales. Quantum sensors, whose performance is governed by quantum rather than thermal fluctuations~\cite{cox2018quantum}, offer a route to transcend these classical barriers. In particular, a frontier lies in reaching the standard quantum limit (SQL) for field detection, where measurement precision is bounded only by quantum projection noise~\cite{caves1981quantum,giovannetti2006quantum,degen2017quantum}.

Rydberg atoms, with their large electric dipole moments and rich transitions, are ideal quantum sensors for electrometry. Initial experiments primarily used thermal vapor cells, probing MW fields via electromagnetically induced transparency (EIT), where a MW-induced Rydberg resonance shift modulates an optical probe transmission, enabling field inference from the ensemble's collective optical response~\cite{sedlacek2012microwave,sedlacek2013atom,holloway2014broadband}. 
Continued advances---including heterodyne~\cite{gordon2019weak,jing2020atomic,prajapati2021enhancement,cai2022sensitivity,jin2025heterodyne}, multi-wave mixing~\cite{kolle2012four,borowka2024continuous,yang2024highly}, multiplexing~\cite{zhang2022rydberg,han2023microwave,cai2023high,knarr2023spatiotemporal,zhang2024ultra,jiao2025arbitrary}, many-body enhancement~\cite{carr2013nonequilibrium,ding2020phase,ding2022enhanced,wu2024nonlinearity,wang2025high} and others~\cite{kumar2017atom,kumar2017rydberg,meyer2021waveguide,cai2023sensitivity,yang2023enhancing,cui2023extending,liang2025cavity}---have steadily improved the performance of vapor-cell-based electrometry.

As research has progressed, however, the challenges of the vapor-cell approach have become apparent.
First, thermal atomic motion limits the coherence time, making it challenging to reach the SQL.
Advances in this direction include MW electrometry using cold-atom ensembles~\cite{han2018coherent,jiao2020single,tu2022high,duverger2024metrology,tu2024approaching}, with recent demonstrations reaching a sensitivity of 2.6 times the SQL~\cite{tu2024approaching}.
Furthermore, the response time is typically limited to the microsecond scale,
arising from atomic population dynamics~\cite{bohaichuk2022origins}.
In addition, the spatial resolution is restricted to the sub-millimeter scale~\cite{holloway2014sub,wade2017real,schlossberger2025two}, set by diffraction-limited optical detection and the large atomic volume required for a measurable signal.

In recent years, programmable atom array has emerged as a powerful system for \emph{in-situ} quantum metrology, enabling remarkable advances in optical clocks, spin squeezing, and magnetometry~\cite{madjarov2019atomic,norcia2019seconds,young2020half,schine2022long,bornet2023scalable,eckner2023realizing,cao2024multi,finkelstein2024universal,schaffner2024quantum}. A key advantage is its direct, site-resolved control and readout of quantum states for individual atomic qubits, bypassing the collective optical response and finite response time of vapor-cell systems. 
This enables a direct route to SQL-level sensitivity and beyond.
Despite these compelling capabilities, its application to high-performance MW electrometry has remained largely unexplored.

Here, we demonstrate a novel MW sensing protocol employing individual Rydberg-level qubits confined in an optical tweezer array.
Each atom operates as an independent, coherent
MW sensor with a field sensitivity only 13\% above the SQL.
The direct coherent evolution of the Rydberg-level qubit yields a 
response time well below \SI{10}{ns},
more than 11 orders of magnitude beyond the Chu limit for a classical antenna of comparable size.
By precisely programming the atom array, we demonstrate \emph{in-situ} mapping of MW near-field distributions with a spatial resolution of $\lambda/3000$, far below the diffraction limit. 
This work establishes the Rydberg-atom array as a powerful platform for MW electrometry, delivering quantum-limited field, temporal, and spatial resolutions.

\emph{Single-atom electrometry with quantum-limited sensitivity}---%
Our experimental protocol, illustrated in Fig.~\ref{Fig:Setup}, divides the tweezer array into two functional regions: a reservoir for atom loading and fluorescence imaging, and a target zone for MW electrometry. Laser-cooled $^{87}$Rb atoms are first loaded into a $15 \times 5$ array of static 808-nm tweezers in the reservoir with a loading probability of $\sim 0.76$. To preserve the long coherence times of the individual atoms by avoiding detrimental dipolar interactions, we shuttle them individually to the target zone rather than exciting the entire array simultaneously to Rydberg states. The trajectories of these mobile tweezers are carefully optimized [Fig.~\ref{Fig:Setup}(a), upper inset] to prevent overlap with the static traps in the reservoir, thereby minimizing cross-talk during transport.

In the target zone, each atom undergoes a sequence of state preparation, Rydberg excitation, MW measurement, and Rydberg-state readout, before being returned for a second imaging, as illustrated in Fig.~\ref{Fig:Setup}(b).
The atom is first prepared in the ground state $\ket{g}=\ket{5S_{1/2},F=2,m_F=2}$ via optical pumping under a magnetic field of \SI{10}{G}.
The Rydberg excitation via stimulated Raman adiabatic passage (STIRAP)~\cite{cubel2005coherent} then transfers the atom from the ground state $\ket{g}$ via $\ket{e}=\ket{5P_{3/2},F=3,m_F=3}$ to a Rydberg state $\ket{\uparrow}=\ket{68D_{5/2},m_J=5/2}$.
After the MW interrogation stage, the Rabi frequency of the signal field resonant with the $\ket{\uparrow}\leftrightarrow\ket{\downarrow}$ transition (where $\ket{\downarrow}=\ket{69P_{3/2},m_J=3/2}$, frequency $f_0=\SI{6.6}{GHz}$) is determined by measuring the population $P_\uparrow$. 
Rydberg-state-selective readout is performed by applying a 480-nm laser pulse resonant with the $\ket{e}\leftrightarrow\ket{\uparrow}$ transition after the MW interaction: this pulse transfers atoms in $\ket{\uparrow}$ back to the ground state for recapture, while atoms remaining in $\ket{\downarrow}$ are lost.
See Appendix~\ref{APD:apparatus} for details.

The MW electric field amplitude $E$ is determined from the measured Rabi frequency $\Omega$ via the relation
$\Omega = E d/\hbar$, where $d$ is the transition dipole moment between the Rydberg states. For the specific states used, the coupling coefficient is $\Omega/E = 2\pi \times \SI{4.88}{kHz/(\micro\volt\,cm^{-1})}$~\cite{vsibalic2017arc}.

For strong signal fields, the field strength can be characterized directly from Rabi oscillations. Figure~\ref{Fig:Setup}(e) shows a representative trace for $\Omega_\mathrm{S} = 2\pi \times \SI{82.6(6)}{\kilo\hertz}$. The absence of significant damping over \SI{20}{\micro\second} demonstrates excellent coherence between the Rydberg levels.
However, in the weak-field regime, Rabi oscillations become unsuitable for precise field characterization. Instead, we employ a two-step measurement sequence: first, a strong local $\pi/2$ pulse ($\Omega_\mathrm{L} = 2\pi \times \SI{1.6}{\mega\hertz}$) prepares the superposition state $\ket{+} = (\ket{\uparrow} + \ket{\downarrow})/\sqrt{2}$, thereby maximizing the population response sensitivity to the subsequent signal field. Then, the superposition state evolves under the signal field.

The population in $\ket{\uparrow}$ after the evolution is related to the signal pulse's Rabi frequency $\Omega_\mathrm{S}$ by
\begin{equation}
    P_\uparrow = \frac{B}{2} - \frac{A}{2}\cos\phi\,\sin(\Omega_{\mathrm{S}}\tau),
\label{eq1}
\end{equation}
where $\tau$ is the signal pulse duration and $\phi$ is the relative phase between the local and signal fields. Here, $A$ and $B$ characterize the state preparation and measurement fidelities and govern the extracted signal and the quantum projection noise. This signal-to-noise ratio (SNR) sets a fundamental floor on the sensitivity for $N$ independent measurements, known as the SQL:
$\sigma_{\Omega}^{\mathrm{SQL}} = \tfrac{\sqrt{B(2-B)}}{A \tau \sqrt{N}}$.
This expression reflects the projection noise of binomial statistics from individual qubit readouts, which is in stark contrast to the Poissonian statistics of ensemble-based measurements. 
This binomial statistics leads to a superior SQL—lower by a factor of $\sqrt{\tfrac{2}{2-B}}$ compared to an ensemble sensor with the same number of atoms. See Appendix~\ref{APD:SQL} for details.

Instead of determining $\Omega_\mathrm{S}$ from $P_\uparrow$ at a fixed phase $\phi$, we scan $\phi$ and measure the peak-to-peak amplitude $\updelta P$ to suppress long-term drifts in the local oscillator strength $\Omega_\mathrm{L}$. The resulting $P_\uparrow$ oscillations are shown in Fig.~\ref{Fig:Setup}(d), and $\updelta P$ relates to the signal Rabi frequency as
$\updelta P = A \sin(\Omega_{\mathrm{S}} \tau)$.
To achieve SQL-level performance, $\updelta P$ is obtained directly from the population difference between $\phi = 0$ and $\pi$.
We refer to this weak-field detection method as the \textit{single-atom homodyne} scheme.

The validity of this approach is confirmed through comparison with an independent power calibration. First, we use a spectrum analyzer to precisely calibrate the MW power fed into the horn antenna across a range of input levels. We then perform Rabi oscillations under these calibrated strong driving fields [Fig.~\ref{Fig:Setup}(e)], which
establishes a direct power-to-$\Omega_\mathrm{S}$ relation [Fig.~\ref{Fig:Setup}(c), cyan squares]. This calibrated relation provides a reference for predicting Rabi frequencies at weaker power levels [Fig.~\ref{Fig:Setup}(c), dashed line]. The excellent agreement between these power-calibrated expectations and the $\Omega_\mathrm{S}$ values obtained via the single-atom homodyne technique in the weak-field regime [Fig.~\ref{Fig:Setup}(c), purple circles] confirms the accuracy of our approach.

\begin{figure}[t]
    \centering
    \includegraphics[width=\columnwidth]{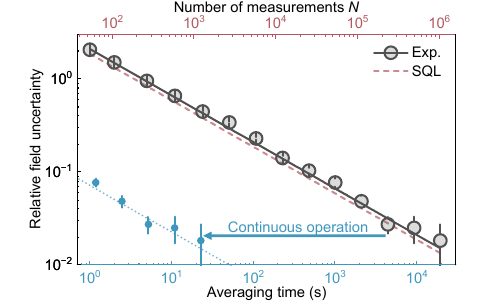}
    \caption{
    Sensitivity approaching the standard quantum limit (SQL). 
    Allan deviation of the measured electric-field amplitude $E=\SI{260}{\nano\volt\per\centi\meter}$ is shown as the relative field uncertainty versus the averaging time $t$ (bottom axis) and the number of measurements $N$ (top axis). 
    A solid line fit ($\propto N^{-1/2}$) to the data yields a single-atom, single-shot sensitivity of $\sigma_E = \SI{3.98(3)}{\micro\volt\per\centi\meter}$. The red dashed line marks the SQL of $\sigma_E^{\mathrm{SQL}} = \SI{3.53(9)}{\micro\volt\per\centi\meter}$.The small difference between $\sigma_E$ and $\sigma_E^{\mathrm{SQL}}$ is analyzed in Appendixes~\ref{APD:SQL} and \ref{APD:cali}.
    The measured sensitivity corresponds to \SI{545(4)}{nV\,cm^{-1}\,Hz^{-1/2}} at the current measurement rate of \SI{53}{\hertz}. 
    The blue dotted line shows the projected sensitivity of \SI{18.7}{nV\,cm^{-1}\,Hz^{-1/2}} achievable at a \SI{45}{\kilo\hertz} rate in the continuous operation mode. 
    Error bars represent $1\sigma$ standard errors.
    }
    \label{Fig:Sensitivity}
\end{figure}

To quantify the sensitivity of our single-atom electrometer in the weak-field regime, we evaluate its single-shot noise-equivalent sensitivity at $E = \SI{260}{\nano\volt\per\cm}$ (Fig.~\ref{Fig:Sensitivity}). 
With a \SI{20}{\micro\second} interaction time per atom, the measured single-atom single-shot sensitivity is $\sigma_E=\SI{3.98(3)}{\micro\volt\per\centi\meter}$, which is only 13\% above the SQL of $\sigma_E^{\mathrm{SQL}} = \SI{3.53(9)}{\micro\volt\per\cm}$ (Appendix~\ref{APD:SQL}).

To characterize the minimum detectable field and long-term measurement stability, we analyze the relative field uncertainty as a function of the number of measurements $N$ (or equivalently, the averaging time), as shown in Fig.~\ref{Fig:Sensitivity}. We obtain a noise-equivalent field resolution of $\sigma_E = \SI{4.7(2.4)}{\nano\volt\per\centi\meter}$ after a total integration time of $\SI{2e4}{\second}$, corresponding to a sensitivity of \SI{545(4)}{\nano\volt\,\cm^{-1}\,\hertz^{-1/2}} at the current measurement rate of \SI{53}{\hertz}. Further improvement in field resolution is anticipated with increased repetition rates. For example, using individual optical addressing and continuous-operation techniques recently developed for atom arrays~\cite{radnaev2025universal,chiu2025continuous}, the repetition rate can be improved to \SI{45}{\kilo\hertz}, yielding a projected sensitivity of \SI{18.7}{\nano\volt\,\cm^{-1}\,\hertz^{-1/2}} (Fig.~\ref{Fig:Sensitivity}, blue dotted line).

Remarkably, the interrogation area of a single-atom sensor ($\sim\SI{2.1e-13}{\meter\squared}$) yields a corresponding power sensitivity of \SI{-211}{dBm/\Hz} at the current measurement rate. This noise-equivalent sensitivity corresponds to the Johnson–Nyquist noise of a classical receiver at an effective temperature of \SI{60}{\milli\kelvin}. At the projected repetition rate of \SI{45}{kHz}, these values improve to \SI{-240}{dBm/\Hz} and \SI{70}{\micro\kelvin}, an effective temperature far below the fundamental cooling limit of any classical electronic receivers.

\emph{Ultrafast temporal response}---%
In addition to achieving standard-quantum-limited sensitivity, our single-qubit scheme enables
ultrafast time-domain MW probing. 
Unlike vapor-cell sensors, whose response time is constrained by finite atomic lifetimes, transit-time effects, and other ensemble-averaging processes, a single-atom qubit responds directly through coherent Hamiltonian dynamics. This enables the faithful detection of MW pulses with negligible distortion on timescales approaching the intrinsic quantum-evolution limit ($\updelta t\propto 1/f_0$).

The response to ultrafast pulses is characterized using the single-atom homodyne scheme, where the sinusoidal oscillation of $P_\uparrow$ is recorded while varying the relative phase $\phi$ between the local and signal fields, and the amplitude $\updelta P$ is extracted from a sinusoidal fit.
In this operation mode, our sensor serves as an atomic vector spectrometer that performs a Fourier transformation and extracts both the amplitude and phase of the signal pulse at the frequency $f_0$ defined by the local field. 
For a rectangular pulse with duration $\tau\gg\updelta t$, the frequency response follows the Fourier spectrum of the pulse, $\updelta P(\updelta f) = A \sin(\Omega_{\mathrm{S}}\tau)\left|\mathrm{sinc}(\updelta f \tau)\right|$, where $\mathrm{sinc}(x) = \sin(\pi x)/(\pi x)$, and $\updelta f$ is the signal frequency offset.

In this case, a 
faithful reconstruction of the Fourier spectrum
indicates an achievable temporal resolution $\updelta t \ll \tau$ and a corresponding bandwidth far beyond $1/\tau$.
Notably, this Fourier transformation is implemented entirely via the coherent evolution of the qubit, inherently free from classical electronic noise (see Appendix~\ref{APD:AFT}).

\begin{figure}[t]
    \includegraphics[width=\columnwidth]{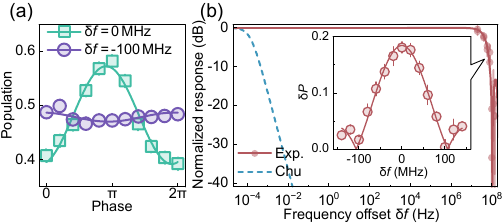}
    \caption{
    Frequency response to an ultrafast pulse.
    (a)~Single-atom homodyne measurements at signal frequency offset $\updelta f=0$ (cyan squares) and \SI{-100}{MHz} (purple circles), with sinusoidal fits.
    (b) The normalized response plotted as a function of $\updelta f$. The measured spectral profile (red solid curve) exhibits a bandwidth exceeding the fundamental Chu limit for a classical antenna of comparable size (blue dashed curve) by 11 orders of magnitude. Inset: Full scan of the frequency response. Each data point represents the fitted amplitude $\updelta P$ obtained from sinusoidal fits as in (a).
    A sinc-type fit (solid curve) to the data yields a pulse duration of \SI{9.4(2)}{\nano\second}, corresponding to a mainlobe width of \SI{213(4)}{MHz}. Error bars represent $1\sigma$ standard errors.
    }
    \label{Fig:Bandwidth}
\end{figure}

We experimentally demonstrate the coherent phase detection of MW pulses as short as $\tau=\SI{10}{\nano\second}$ [Fig.~\ref{Fig:Bandwidth}(a)], with a strong Rabi coupling of $\Omega_\mathrm{S}=2\pi\times\SI{6.37}{\mega\hertz}$ that maintains a high signal-to-noise ratio. 
To suppress fine-structure mixing at such high drive strengths, the magnetic field defining the quantization axis is raised to $\sim\SI{30}{G}$, providing a Zeeman splitting of $\sim 2\pi\times \SI{50}{\mega\hertz}$ between neighboring states.
The measured frequency response [Fig.~\ref{Fig:Bandwidth}(b)] displays a 
mainlobe width of \SI{213(4)}{MHz},
which lies well within the intrinsic bandwidth of the single-atom sensor, corresponding to the ability to track nanosecond-level MW transients.

The results demonstrate the capability of quantum sensors to surpass the Chu limit~\cite{chu1948physical}, which fundamentally constrains the temporal response and bandwidth of any passive classical antenna.
For the single $68D_{5/2}$ Rydberg atom used here, the mean orbital radius is $\expval{r}\approx\SI{260}{\nm}$, significantly smaller than the wavelength of $\lambda=\SI{45}{mm}$.
The corresponding Chu-limited field-response FWHM bandwidth for a classical antenna of comparable size is only
$\mathrm{BW}_\mathrm{Chu} = f_0\left(\tfrac{2\pi\expval{r}}{\lambda}\right)^3 \approx \SI{0.3}{\milli\hertz}$.
The atomic intrinsic bandwidth $\gg\SI{213}{\mega\hertz}$ exceeds this classical bound by more than 11 orders of magnitude.

\begin{figure}[t]
    \includegraphics[width=\columnwidth]{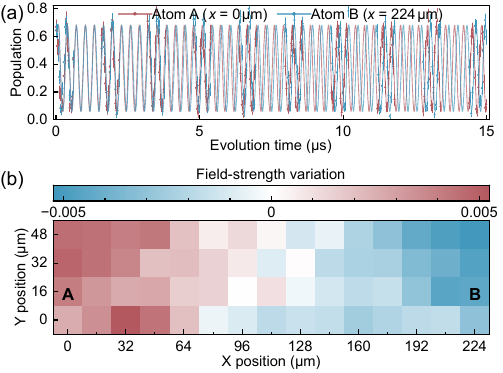}
    \caption{
    \emph{In-situ} measurement of sub-wavelength field distribution.
    (a) Rabi oscillations recorded via the population in $\ket{\uparrow}$. Sinusoidal fits yield Rabi frequencies of $\Omega_\mathrm{A} = 2\pi \times \SI{3.187(1)}{\mega\hertz}$ and $\Omega_\mathrm{B} = 2\pi \times \SI{3.160(1)}{\mega\hertz}$ for atom A ($x = \SI{0}{\micro\meter}$) and atom B ($x = \SI{224}{\micro\meter}$), respectively. The oscillations accumulate a measurable phase shift of approximately 0.4 periods over 47 Rabi cycles.
    (b) Spatial map of the relative field strength. The local Rabi frequency at each position $(x, y)$ is determined from a sinusoidal fit to the continuous Rabi oscillation. The color scale represents the normalized field variation $\zeta(x, y)$.
    }
    \label{Fig:In-situ}
\end{figure}

\emph{Sub-wavelength in-situ electrometry}---%
Apart from its temporal resolution, our atomic sensor also breaks the spatial-resolution barrier that constrains conventional MW antennas.
While achieving sub-wavelength spatial resolution is fundamentally difficult for classical receivers, atom-based MW sensors provide a new avenue for high-resolution MW imaging, since the matter--field interaction region can be engineered to be much smaller than the MW wavelength. 
In vapor-cell-based Rydberg electrometry, spatial resolution on the \SI{0.1}{mm} scale~\cite{holloway2014sub,wade2017real,schlossberger2025two} has been achieved, constrained by the laser-beam waists and optical path lengths that support sufficient optical densities. 
In contrast, our single-atom sensor utilizes an interaction region defined by the Rydberg electron wavefunction, with a mean radius of $\langle r \rangle \sim \SI{260}{\nano\meter}$, enabling sub-micrometer spatial resolution.

Resolving field distributions at the micrometer scale demands both high spatial resolution and the ability to detect minute variations in field strength. To generate a measurable field gradient at this scale, sharp-edged metallic structures are positioned approximately $\lambda/2$ away along the $\pm$x direction, enhancing the near-field inhomogeneity. Despite this engineered configuration, the field variation across micrometer distances remains extremely small.

To map such fine spatial variations, resonant Rabi driving is employed. Figure~\ref{Fig:In-situ}(a) shows continuous Rabi oscillations for two atoms (A and B) separated by \SI{224}{\micro\meter}. The oscillation contrast remains nearly constant over \SI{15}{\micro\second} (corresponding to 47 full $2\pi$ cycles),  demonstrating excellent coherence of the Rydberg-level qubits with no observable damping. Although the MW field difference between the atoms results in only a slight variation in Rabi frequency $\Omega$, the accumulated phase difference during extended driving produces a clearly detectable oscillation mismatch. 
The local field-strength variation for the $i$-th atom, $\zeta_i$, is obtained as $\zeta_i=(|\Omega_i|-\bar{|\Omega|})/\bar{|\Omega|}$, since $\Omega\propto E$. See Appendix~\ref{APD:insitu} for details.

Based on this approach, the atomic position is scanned across the target zone to reconstruct the full field distribution $\zeta(x,y)$ [Fig.~\ref{Fig:In-situ}(b)]. The local Rabi frequency at each lattice site is extracted from a sinusoidal fit to the observed Rabi oscillations. A scanning step of \SI{16}{\micro\meter} is sufficient to resolve the spatial features of the field, with variations between adjacent pixels less than $0.1\%$.

\emph{Summary and outlook}---%
In summary, our work establishes Rydberg-atom arrays as a coherent platform that delivers simultaneous quantum-limited performance across field, temporal, and spatial resolutions
on a single, microscopic device.
The current sensitivity can be further improved via continuous-operation protocols already demonstrated in atom-array architectures. Looking forward, the system provides a clear pathway to surpass the SQL by generating many-body entangled states via programmable Rydberg interactions~\cite{pezze2018quantum,facon2016sensitive,gilmore2021quantum}.
Furthermore, our sensor's performance signifies an advance in temporal fidelity, not merely spectral coverage. It captures the waveform of fast pulses, a task distinct from, and more demanding than, the frequency-range coverage offered by the multiplexing techniques in vapor-cell sensors.

Beyond serving as a benchmark for quantum-limited electrometry, this platform opens avenues for a range of applications. 
These include high-resolution near-field mapping of integrated photonic and MW circuits, capturing short, information-rich MW transients within the framework of Shannon-limited communication~\cite{Shannon1948}, functioning as a quantum receiver capable of detecting ultra-weak signals and, specifically, operating as an atomic vector spectrometer for extracting both the amplitude and phase of MW fields. This platform's inherent immunity to classical electronic noise further enhances its potential for fundamental research, such as the detection of faint electromagnetic signatures potentially associated with dark matter~\cite{bradley2003microwave,sushkov2023quantum,graham2024rydberg}.

\nolinenumbers

\emph{Acknowledgements}---%
We thank Linjie Zhang, Binbin Wei, Mingyong Jing, Yuan Sun, Chen Zhou, Hongping Liu, Weibin Li, Dong-Ling Deng, Dong Yuan and Yiqiu Ma for insightful discussions. This work was supported by the National Key Research and Development Program of China (Grant No.~2021YFA1402003), the National Science and Technology Major Project of the Ministry of Science and Technology of China (Grant No.~2023ZD0300901), the National Natural Science Foundation of China (Grant Nos.~12374329, U21A6006 and 12404580), the Science and Technology Commission of Shanghai Municipality (Grant No.~25LZ2601002). Y.Z. acknowledges support from the Shanghai Qi Zhi Institute Innovation Program SQZ202317.

\nolinenumbers
\bibliography{ref}

\clearpage
\appendix
\appendixsection{Experimental apparatus}
\label{APD:apparatus}
The experiment builds on our previous setup reported in Ref.~\cite{Xiang2024Observation}. 
Here we summarize only the elements directly relevant to the present work. 
A more detailed description of the optical layout and cooling sequence can 
be found in the Supplementary Information of Ref.~\cite{Xiang2024Observation}.

\emph{Array preparation and atom shuttling}---%
The experiment uses single $^{87}$Rb atoms trapped in optical tweezers, with a movable tweezer that transfers atoms from a reservoir to the target zone. Individual atoms are first loaded into static traps in the reservoir via gray-molasses light-assisted collisions, achieving a single-atom loading probability of 0.7591(2) as verified by fluorescence imaging. The loaded atoms are then sequentially shuttled to the target zone.
Rather than computing optimized trajectories that skip empty traps in real time, we employ a fixed preset path for the moving tweezer that treats all traps uniformly [Fig.~\ref{Fig:Setup}(a), upper inset]. This choice simplifies hardware control and avoids significant overhead in waveform generation and input/output handling, resulting in faster overall transport despite the occasional extra moves.

\emph{Microwave control}---%
Precise control of the amplitude and relative phase of the local and signal microwave (MW) fields is implemented using two channels of an arbitrary waveform generator (AWG, M4i.6631-x8, Spectrum Instrumentation). The two outputs are combined with a power splitter (ZFSC-2-1+, Mini-Circuits) and mixed with an MW source (SMB100A, Rohde \& Schwarz) using an IQ mixer (MMIQ-1040L, Marki) to generate the required MW pulses. The AWG software determines the pulse envelopes, timing, and inter-channel phase.
To suppress quantization artifacts when generating weak signal pulses, we insert additional attenuation on the corresponding AWG channel: \SI{40}{dB} for Rabi frequencies above $2\pi\times\SI{10}{kHz}$ (power range: \SI{-51}{dBm} to \SI{-75}{dBm}), and \SI{70}{dB} when the Rabi frequency is below $2\pi\times\SI{10}{kHz}$ (power range: \SI{-81}{dBm} to \SI{-105}{dBm}).

\emph{State-selective detection}---%
The Rydberg-state-selective readout is performed using a two-step protocol. Atoms remaining in $\ket{\uparrow}$ are first transferred back to $\ket{g}$ using a resonant 480-nm pulse. Atoms in $\ket{\downarrow}$ are then removed by switching the repulsive optical-tweezer potential back on, which is otherwise held off during Rydberg operations.

\appendixsection{Analysis of standard quantum limit}
\label{APD:SQL}
Our single-atom MW sensor achieves quantum-limited performance across multiple dimensions: its field sensitivity is limited only by the fundamental quantum projection noise (QPN), temporal resolution by the energy splitting of the qubit levels ($\updelta t \propto 1/f_0$), and spatial resolution by the atomic wavefunction distribution ($\updelta x \sim \expval{r}$).
This intrinsic quantum-limited nature sets our sensor apart from classical MW probes, as well as from ensemble-based Rydberg sensors, whose collective measurements are fundamentally constrained by additional statistical averaging, finite response time and large volume required to maintain sufficient optical depth.

To quantify the ultimate sensitivity of our single-atom device, we now analyze the standard quantum limit (SQL) arising from the QPN.
According to Eq.~(\ref{eq1}), for small $\Omega_{\mathrm{S}}$, the optimal sensitivity is quantified by the Fisher information $F$ under conditions $\phi = 0$ or $\pi$,
\begin{equation}
    F = N \frac{A^2 \tau^2}{B(2-B)},
\end{equation}
where $N$ is the number of independent measurements. 
Unlike ensemble-based approaches that treat all $N$ atoms as a single collective Bloch vector of length $J = \tfrac{AN}{2}$ and contrast $C = \tfrac{A}{B}$---with QPN described by Poisson statistics and signal-to-noise ratio $\sqrt{JC}$---the individual readout of each qubit resolves the full binomial statistics of the measured populations, with variance $\sigma_P^2 = \tfrac{1}{4} B(2-B)$, thereby extracting the maximum information allowed by quantum mechanics. This yields the SQL for the sensitivity,
\begin{equation}
    \sigma_{\Omega}^{\mathrm{SQL}} = \frac{1}{\sqrt{F}} = \frac{\sqrt{B(2-B)}}{A \tau \sqrt{N}},
\end{equation}
as set by the Cram{\'e}r--Rao bound, corresponding to a $\sqrt{\tfrac{2}{2-B}}$-fold improvement over an ensemble-based sensor with the same atom number and field response but a Poissonian distribution, which is approximated by a worst-case binomial scenario where $A,B\to0$, as the information about the presence of any qubit involved in the measurement is totally lost. This perspective aligns with the analysis of the SQL baseline in existing work on quantum metrology using individually resolved atoms (e.g., Ref.~\cite{eckner2023realizing}), emphasizing that our single-atom sensor not only operates closer to the SQL but also improves the SQL baseline itself.

The parameters $A = 0.438(11)$ and $B = 0.680(6)$ are determined from sinusoidal fits of Rabi oscillations [Fig.~\ref{Fig:Setup}(e)], which yield $\sigma_{\Omega}^{\mathrm{SQL}}=2\pi\times\SI{17.2(4)}{kHz}/\sqrt{N}$ and $\sigma_E^{\mathrm{SQL}} = \SI{3.53(9)}{\micro\volt\per\cm}/\sqrt{N}$. 
The measured sensitivity falls slightly short of the SQL due to the finite coherent time [$T_2\approx\SI{60}{\us}$ given by the damping-sine fit in Fig.~\ref{Fig:Setup}(e)] and system instability (as described in Appendix~\ref{APD:cali}).
As the drifts in $A$ result in a varying baseline of the SQL, we use the maximum fitted value of $A=0.438(11)$ from all recorded data when evaluating the SQL to adopt the most conservative standard. Even under this stringent standard, the measured noise-equivalent field sensitivity is still only 13\% above the SQL. 

\appendixsection{Calibration details}
\label{APD:cali}
We perform real-time calibration of multiple experimental parameters to improve long-term stability of the MW measurements. The Rabi oscillation amplitude, $A$, is sensitive to the atomic temperature, which affects the recapture--loss rate in the target zone. This temperature slowly drifts over time due to misalignment between the dynamic and static tweezers, as the atom experiences heating during poorly aligned transfers. To address this, one column of the $15\times5$ array is programmed to undergo a modified sequence that omits all Rydberg operations. The survival probability of these atoms, $R$, is used to calibrate the measured $\updelta P$ via $\updelta P_\mathrm{cal} = \updelta P_\mathrm{raw}/R$.

In addition to the amplitude calibration described above, we also calibrated the MW resonance frequency and the relative phase between the local and signal fields, although the measurement outcome depends on these two parameters only to the second order, unlike the first-order sensitivity to $A$. Slow drifts of the background electric field, induced by the dynamic distribution of surface charges on the vacuum cell, cause small shifts of the Rydberg levels and thus an effective detuning of the MW resonance. To compensate for this effect, we record MW Ramsey spectra on a daily basis,
extract the detuning from the fits, and apply a bias field to maintain resonance.
Furthermore, slow thermal drifts in the MW circuitry and cables lead to slight variations in the relative phase between the local and signal fields. We calibrate this phase by scanning a full single-atom homodyne oscillation and fitting the software-defined phase corresponding to $\phi=0$ and $\pi$. This calibration is repeated 
on a daily basis,
and the updated phase offsets are applied in the control software.

\begin{figure}[t]
    \includegraphics[width=\columnwidth]{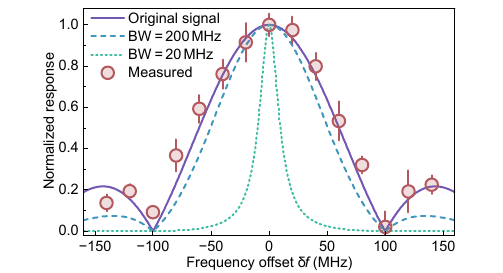}
    \caption{Ultrafast pulse spectra reconstructed by receivers with different bandwidths. The 10-ns signal pulse exhibits a sinc-type spectrum with mainlobe width \SI{200}{MHz} (purple solid curve), which is filtered by sensors with comparable (blue dashed curve) or significantly lower bandwidth (cyan dotted curve). The measured data fall close to the spectrum of the original signal, indicating a bandwidth of $\gg\SI{200}{MHz}$.}
    \label{Fig:zhaobu}
\end{figure}

\appendixsection{Atomic Fourier transformation}
\label{APD:AFT}
To probe the temporal resolution of the single-qubit sensor, a rectangular signal pulse with a duration of $\tau=\SI{10}{ns}$ (limited by the nanosecond-level rise/fall time of MW switches) is generated with different frequency offset $\updelta f$ and initial phase $\phi$ relevant to the local field. The atomic homodyne scheme is equivalent to a Fourier transformation that extracts both the amplitude and phase $\phi$ at a given frequency offset $\updelta f$, as the evolution operator for a small signal is
$U(\tau) = \mathbb{I}-\frac{i}{\hbar}\int_0^{\tau} H(t)\mathrm{d}t$
as a direct result of the Schr\"odinger equation.
Here, the Hamiltonian $H(t)=\hbar\Omega(t) e^{i(\phi-2\pi\updelta ft)}\ket{1}\bra{0}+\mathrm{H.c.}$ contains the kernel $e^{-i2\pi\updelta ft}$ of the Fourier transformation. The rotation of Bloch vector
$\updelta \hat{\sigma}\propto\int_0^{\tau} \Omega(t) e^{i(\phi-2\pi\updelta ft)} \mathrm{d}t$
is thus directly connected to the Fourier transformation $F(\updelta f) = \int_0^\tau E(t) e^{i(\phi-2\pi\updelta ft)} \mathrm{d}t$ of an MW pulse with a time-dependent electric field $E(t)\propto \Omega(t)$. Therefore, the homodyne detection extracts the Fourier spectrum of the signal pulse as long as the response time (bandwidth) of the atomic sensor is faster (larger) than the signal's duration (frequency-domain spreading).

The frequency response can be alternatively understood in the time domain. 
For a narrow-band sensor that distorts and stretches the pulse, a small $\updelta f$ comparable to its bandwidth causes significant phase mismatch over an extended pulse duration, and the homodyne results $\updelta P$ are canceled out if the pulse is stretched to as long as $1/\updelta f$. Therefore, the reconstructed signal spectrum is narrowed (Fig.~\ref{Fig:zhaobu}, cyan curve). In contrast, a sensor with a temporal resolution comparable to or far better than \SI{10}{ns} can capture the phase until $\updelta f$ approaches $1/\tau$, and exhibit a frequency response nearly identical to the Fourier spectrum of the signal pulse, i.e., a sinc function with characteristic zeros at $\updelta f=\pm n/\tau, n=1,2,\dots$ and sidelobes between them (Fig.~\ref{Fig:zhaobu}, blue and purple curves).

\appendixsection{In-situ measurement details}
\label{APD:insitu}
In Fig.~\ref{Fig:In-situ}(a), the Rabi oscillations show a slight acceleration before stabilizing after approximately \SI{5}{\micro\second}. This behavior is attributed to the finite time required for the MW circuitry to reach a steady operating state after the pulse is turned on. Accordingly, when fitting the data with a sinusoidal function, the weights for the data points within the first \SI{5}{\micro\second} are set to zero, yielding an effective onset time of $t_0 \approx \SI{0.08}{\micro\second}$. The same value of $t_0$ is used consistently in the analysis of the Rabi oscillations presented in Fig.~\ref{Fig:In-situ}(b), where the local Rabi frequency at each site is determined by sinusoidal fits of the last period of the observed Rabi oscillations with $t_0$ fixed at \SI{0.08}{\micro\second} and Rabi frequency $\Omega$ constrained within the range of $2\pi\times\SI{3.16}{MHz}\pm1\%$.
The \emph{in-situ} measurements, which aim to resolve field variations at the 0.1\% level, are sensitive to both the MW source power drifts and the spatial variations of the near-field pattern. Although the Rabi oscillations for two atoms in the same row shown in Fig.~\ref{Fig:In-situ}(a) are recorded simultaneously, the data supporting Fig.~\ref{Fig:In-situ}(b) are acquired sequentially: measurements for each row are completed before moving to the next row. 

\end{document}